\documentclass[a4paper]{article}

\usepackage{adjustbox}
\usepackage{hyperref}
\usepackage{booktabs}
\usepackage{graphicx}
\usepackage{float}
\usepackage{tabto}
\usepackage{subcaption}
\usepackage{INTERSPEECH2020}
\usepackage[dvipsnames]{xcolor}
\usepackage{bm}
\input{def.set}

\mathchardef\mhyphen="2D % Define a "math hyphen"

\title{
Sparse Mixture of Local Experts for Efficient Speech Enhancement}
\name{
Aswin Sivaraman, Minje Kim}
\address{
Indiana University, Department of Intelligent Systems Engineering, USA}
\email{
asivara@indiana.edu, minje@indiana.edu}

\begin{document}

\maketitle

\begin{abstract}
In this paper, we investigate a deep learning approach for speech denoising through an efficient ensemble of specialist neural networks. By splitting up the speech denoising task into non-overlapping subproblems and introducing a classifier, we are able to improve denoising performance while also reducing computational complexity. More specifically, the proposed model incorporates a gating network which assigns noisy speech signals to an appropriate specialist network based on either speech degradation level or speaker gender. In our experiments, a baseline recurrent network is compared against an ensemble of similarly-designed smaller recurrent networks regulated by the auxiliary gating network. Using stochastically generated batches from a large noisy speech corpus, the proposed model learns to estimate a time-frequency masking matrix based on the magnitude spectrogram of an input mixture signal. Both baseline and specialist networks are trained to estimate the ideal ratio mask, while the gating network is trained to perform subproblem classification. Our findings demonstrate that a fine-tuned ensemble network is able to exceed the speech denoising capabilities of a generalist network, doing so with fewer model parameters. \\
\end{abstract}


\section{Introduction}

Speech denoising is a highly-studied, wide-ranging research problem aimed at improving speech quality and intelligibility within noisy recordings. Methods to this end are often assessed by the removal of non-speech components and the minimization of any artifacts introduced by the enhancement process. In this work, we address the specific scenario of removing non-stationary uncorrelated background noise from a monophonic recording of a single English speaker. While there are well-established algorithms for speech denoising---such as Wiener filtering \cite{WienerN1964}, spectral subtraction \cite{BollSF79ieeeassp}, and the short-time spectral amplitude method \cite{EphraimY1984spectralamplitude}---recent advances in deep learning technology have significantly improved performance by reformulating speech denoising as a supervised learning task.

Conventionally, deep neural networks (DNN) address speech denoising either by directly estimating the clean signal or by estimating a mask. Multilayer perceptrons, or densely-connected neural networks, were first used to perform binary classification between speech and noise (i.e. by masking in the time-frequency domain) \cite{WangY2013ieeeaslp, XuY2014ieeespl}. To model the temporal structure inherent in speech signals, more complex deep learning technologies such as recurrent neural networks (RNN) have seen greater usage \cite{WeningerF2015lvaica, SunL2017hscma}.

Masking algorithms often require explicit knowledge about the exact number of sources. However, a recent DNN-based approach known as ``deep clustering" performs speaker-independent separation on an arbitrary number of sources by estimating spectrogram embeddings \cite{HersheyJ2016icassp}. Deep clustering has been shown to benefit from jointly estimating time-frequency masks \cite{LuoY2017chimeranet} or through incorporating spatial audio features such as phase difference \cite{TzinisE2019icassp}.

The growing number of DNN-based speech denoising methods is a consequence of the ubiquitous increase in computing power, in part due to accelerated matrix multiplications on graphics processing units (GPUs). In order to learn highly non-linear mapping functions, modern-day neural networks now operate on the scale of millions of learnable parameters \cite{KrizhevskyA2012alexnet}. However, this trend conflicts with the commercial demand for robust low-power models, designed for deployment on embedded systems or resource-limited devices. Hence, the fundamental trade-off between model performance and model complexity is the subject of ongoing deep learning research. Popular domain-agnostic techniques for network compression include pruning weights and filters \cite{HanS2016iclr, LiH2016pruning, LuoJH2017thinet} and quantizing network parameters \cite{SoudryD2014nips, RastegariMCoRR16, KimMJ2015icmlw}. With regards to speech denoising, bit-depth reduction has been successfully used to compress fully-connected and recurrent models \cite{KimMJ2018icassp, KimSW2019icassp}.

This paper examines an approach to network compression through structural modification to an ordinary network architecture. Unlike pruning methods---which thin out inactive components from a full topology---our approach considers a model made up of independent sub-modules, each of which specializes in a particular subproblem. This neural network design philosophy, popularly known as ``mixture of local experts" (MLE), has seen widespread study \cite{JacobsR1990nips}. Within speech enhancement or source separation literature, MLE architectures have been introduced to achieve a more adaptive ensemble of specialist modules \cite{KimMJ2017icassp, ChazanSE2017MoERNN, LiuA2019selection}. However, their implication in the sense of network compression has not been largely explored, which is the main contribution of this paper. 

Recent research has shown that the speech denoising problem can be decomposed into independent subproblems which constitute the various dimensions along which noisy speech signals may vary \cite{KolbakM2017ieeeacmaslp}. We develop this intuition further within an MLE model, whose classifier sub-module must choose only the most relevant specialist to process a given test sample. Because the inference is done only on the chosen specialist, which possesses a smaller number of parameters than an equally-performing generalist model, the overall computational complexity is reduced. Our experiments comprehensively evaluate the sparsely active ensemble of specialists architecture, showing that the reduced model complexity does not compromise speech denoising performance. 

\begin{figure}[ht]
    \centering
    \begin{subfigure}[t]{.66\columnwidth}
        \centering
        % [height=1.45in]
        \includegraphics[width=\columnwidth]{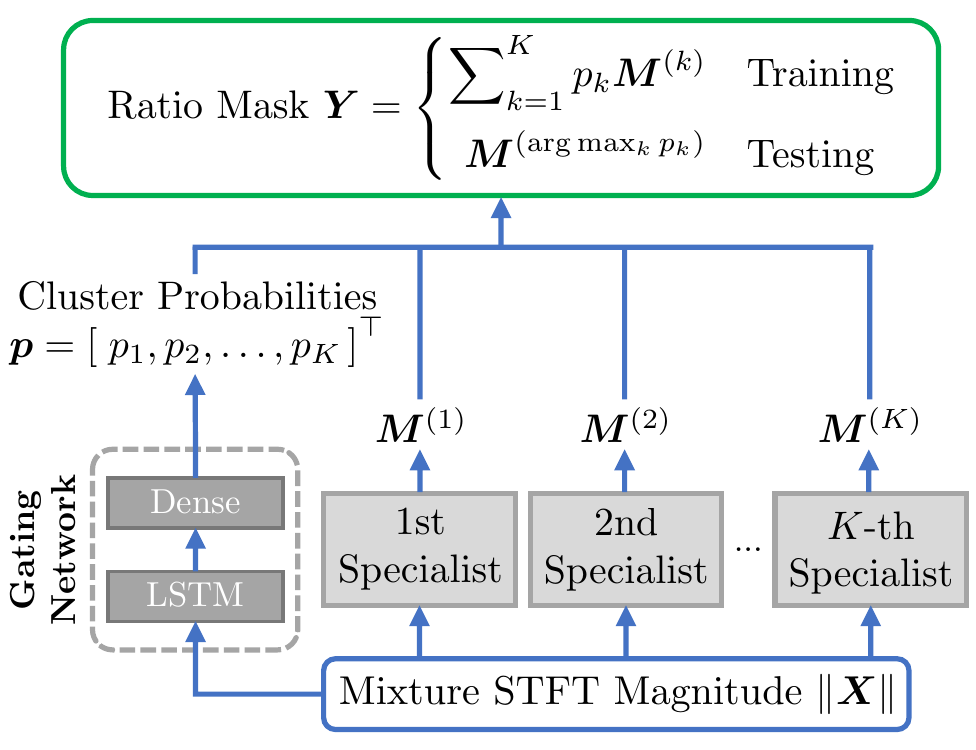}
        \caption{}
        \label{fig:arch_ensemble}
    \end{subfigure}
    \hfill
    \begin{subfigure}[t]{.33\columnwidth}
        \centering
        % [height=1.45in]
        \includegraphics[width=\columnwidth]{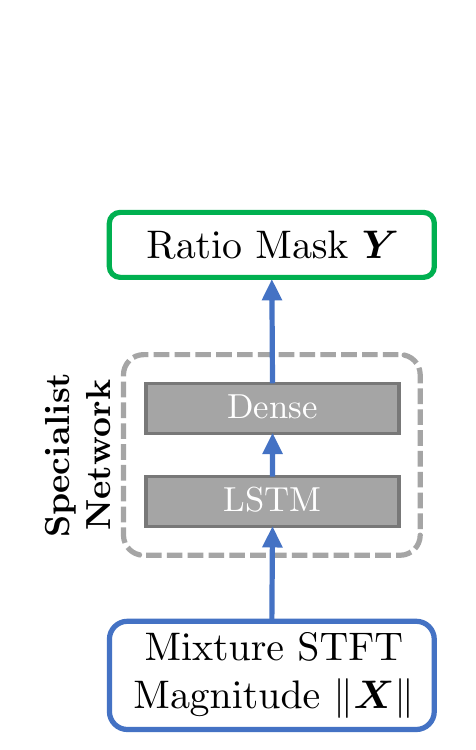}
        \caption{}
        \label{fig:arch_baseline}
    \end{subfigure}%
    \caption{Comparison between (a) the proposed ensemble of specialists model and  (b) the baseline model.}
    \label{fig:architecture}
\end{figure}

\section{The Proposed Method}

Given that the speech denoising task can be divided into mutually exclusive subproblems, we propose that it must be possible to split a complete noisy speech dataset along some latent dimension in order to form non-overlapping subsets (i.e. clusters). Although the MLE architecture is theoretically capable of learning latent clusters in an unsupervised fashion, in this work, we utilize prior knowledge about the problem domain to designate two latent spaces: (1) different speech degradation levels and (2) speaker gender.

The proposed model, shown in Figure \ref{fig:arch_ensemble}, is an ensemble of specialist networks regulated by a gating network. While it is fundamentally possible to utilize the inferences of multiple specialists, we propose using only a single specialist in order to bring computational complexity during inference to a minimum. We assume that the noisy speech data can be split into distinct subsets. Consequently, we pre-train each specialist network to individually address one subproblem. Our experiments compare the proposed ensemble model against a baseline model, shown in Figure \ref{fig:arch_baseline}, which is architecturally equivalent to a specialist network but is trained using the entire noisy speech training set. In the following subsections of this paper, we define the specialist and gating modules more formally.

\subsection{Specialist Networks}

A monaural time-domain mixture signal $\bm{x}$ is defined as the sum of a clean speech signal $\bm{s}$ and an additive background noise signal $\bm{n}$: $\bm{x} = \bm{s} + \bm{n}$. The goal of speech denoising is to learn a mapping function $g$ which produces an estimated signal $\bm{\hat{s}}$ such that $g(\bm{x}) = \bm{\hat{s}} \approx \bm{s}$.

A well-known objective metric for this single-channel denoising task is the signal-to-distortion ratio (SDR). Implemented as part of the \textit{BSS\_eval} toolkit \cite{VincentE2006ieeeaslp}, SDR expresses the ratio of energy between a source signal and an estimate signal. A more robust modification of SDR, known as scale-invariant SDR (SI-SDR), uses a scaling factor $\alpha$ to ensure that the residual vector (between source $\bs$ and estimate $\hat{\bs}$) maintains orthogonality to the source \cite{LeRouxJL2018sisdr} as follows:

\begin{equation}
\operatorname{SI-SDR} (\bs, \hat{\bs}) 
=
%\limits_{j}
10 \log_{10}\left[ \frac{\sum_t (\alpha s_t)^{2} }{\sum_t (\alpha s_t - \hat{s}_t )^{2} } \right].
\label{eq:sisdr}
\end{equation}

For standard SDR, $\alpha = 1$; for SI-SDR, $\alpha = \frac{ \hat{\bs} ^\top \bs }{ \bs^\top\bs }$. Both specialist and baseline networks are trained to maximize this metric between the recovered estimate speech $\hat{\bs}$ and the reference clean speech $\bm{s}$.

There are many possible ways to produce $\hat{\bs}$ given only $\bm{x}$. One established approach is known as time-frequency (TF) masking, in which models estimate a TF mask matrix $\bm{Y}$ such that $\hat{\bS} = \bm{Y} \odot  \bm{X}$, where $\odot$ denotes Hadamard product and $\hat{\bS}$ and $\bm{X}$ are the discrete short-time Fourier transforms (STFT) of the estimate signal and the noisy mixture signal respectively. The mask matrix is a ratio at each TF-point in the mixture signal belonging to either noise or speech, with values between 0 and 1 respectively. The inverse STFT transforms $\hat{\bS}$ from the time-frequency domain back to the time domain $\hat{\bs}$. To estimate $\bm{Y}$ through supervised learning, both specialist and baseline models target the ideal ratio mask (IRM) \cite{SrinivasanS2006irm}, which is defined as:

\begin{equation}
\operatorname{IRM} =
\sqrt{
\frac{ |\bm{S}| ^{2} }{ |\bm{S}| ^{2} + |\bm{N}| ^{2} }
}    
\end{equation}

$|\bm{S}|$ and $|\bm{N}|$ denote the magnitude STFT of speech and noise respectively. IRM has been shown to work well as a masking target assuming that the interfering noise signal $\bm{n}$ is uncorrelated with target speech signal $\bm{s}$ \cite{NarayananA2013icassp, WangY2014transactions}.

To focus our attention on the benefits of the ensemble philosophy, with consideration for the constraints of resource-limited environments, we design our specialist network with unidirectional recurrent layers followed by a feed-forward dense layer. The recurrent layers are made up of long short-term memory (LSTM) cells  \cite{HochreiterS1997nc}. The number of recurrent layers as well as the number of hidden units per layer are adjustable experiment parameters which affect the overall complexity of the model. The specialist network takes the noisy speech magnitude STFT $\bm{|X|}$ as input and predicts a ratio mask matrix $\bm{Y}$. Subsequently, $\operatorname{inv-STFT} \left( \bm{Y} \odot \bm{X} \right)$ yields the denoised speech estimate $\hat{\bs}$.

We note that convolutional neural networks (CNN) on time-domain signals currently achieve the state-of-the-art performance in source separation \cite{LuoY2019conv-tasnet}. Despite their low model complexity, convolutional architectures are able learn the sequence-to-sequence mapping. We leave general application of our proposed ensemble model to different architectures for future work.

\subsection{Gating Network}
 
The gating network is responsible for assigning an input signal to the appropriate specialist. It introduces a classification sub-task as overhead to the overarching denoising task, splitting the full training dataset into some number of latent clusters.

Identifying latent clusters in a noisy speech corpus is non-trivial. Prior works using ensemble models for speech enhancement have shown that specialists may be trained to denoise a particular phoneme \cite{ChazanSE2017MoERNN}. This approach, which requires training data to be phoneme-labeled, is naturally language-dependent but also non-sparse, as multiple specialists may actively perform some computations due to the high variance of phonemes in speech.
% Note: the model could theoretically learn some weird speaker-based specialists someday
% Designing specialists which are speaker-specific or noise type-specific could potentially limit the model's performance on data outside of the training set.
To ensure a sparse activation of specialists (ideally one specialist per input signal), a more generalized latent clustering is preferred. For this reason, we design two types of gating networks to classify inputs based on either \textit{speech degradation level} or \textit{speaker gender}.

Similar to the specialist architecture, our gating networks are also designed with multiple recurrent layers and a single dense layer. However, in our current proposed model the gating network does not make predictions frame-by-frame; after processing the entire input sequence, the network produces a single softmax vector $\bp$, with $K$ elements corresponding to the number of clusters (i.e. the number of specialists). The index of the maximum value in $\bp$ should correspond to the index of the best-suited specialist.

\subsection{Ensemble Network}

The proposed ensemble model combines $K$ specialist networks together with a gating network. First, all of the sub-networks are independently trained. The combination of these pre-trained modules forms a primitive ensemble, as the gating network can already assign an incoming test example to one of the specialists. The output mask $\bY$ is chosen from the specialist which corresponds to the maximum value of gating network softmax vector $\bp$. The ``hard" gating mechanism is formulated as:

\begin{equation}
    \bY = \bM^{(k^*)}, \quad k^*=\argmax_k p_k,
\label{eq:hard_ensemble}
\end{equation}

where $\bM^{(k)}$ denotes the predicted ratio mask matrix from the $k$-th specialist.

However, this na\"ive ensemble is sub-optimal as it lacks the potential co-adaptation between gating and specialist networks. For example, given the fact that the gating network cannot classify mixtures with 100\% accuracy, the specialists should adapt to the situation where it processes a misclassified sample (e.g., a male speech sample falls in the female speaker's specialist). Knowing this, we can further train the sub-modules in unison. During this fine-tuning phase, the ensemble model estimates the output ratio mask $\bm{Y}$ by performing a normalized sum over the individual masks $\bm{M}^{(k)}$ produced by all specialists weighted by the gating network softmax vector $\bp$. This ``soft" gating mechanism ensures that the ratio mask calculation is differentiable, and is formulated as: 

\begin{equation}
\bY = \sum_k p_k\bM^{(k)}.
\label{eq:soft_ensemble}
\end{equation}

During the test phase, the weighted sum is replaced by the hard-decision shown in Eq. \ref{eq:hard_ensemble}. This difference between training-time and evaluation-time computation in the ensemble architecture is the crux of its efficiency; only one out of all the specialists is used to process the entire mixture spectrogram $\bm{|X|}$, making the total used network parameters a fraction of the total learned. We reduce the discrepancy between the hard and soft gating mechanisms, used during testing and fine-tuning respectively, by introducing a scaling parameter $\lambda$ to the softmax gating network output:

\begin{equation}
p_k = \frac{\operatorname{exp}(\lambda \cdot o_k)}
{\sum_{j=1}^{K} \operatorname{exp}(\lambda \cdot o_j)}.
\end{equation}

Each element of the gating network output cluster probability vector ($p_k$) is dependent on the corresponding element of dense layer output ($o_k$) normalized by the sum of all dense layer output elements. While the traditional softmax function can be calculated using $\lambda = 1$, we elevate the sparsity of $\bp$ by setting $\lambda = 10$. This saturates $\bp$ to be near-1 at a single index and near-0 at every other index, making the weighted sum for ratio mask $\bm{Y}$ (Eq. \ref{eq:soft_ensemble}) effectively select the best-case specialist mask. This modification of the softmax function has been successfully used for quantizing vectors with image compression \cite{AgustssonE2017softmax} and for speech coding \cite{KankanahalliS2018icassp}.

\section{Experiment Setup}

All models (specialist, gating, baseline, and ensemble) are trained using a stochastic data sampling strategy which dynamically mixes clean speech recordings from the LibriSpeech\footnote{Available for download at \url{http://www.openslr.org/12/}.} corpus \cite{PanayotovV2015Librispeech} with noise recordings from the  MUSAN\footnote{Available for download at \url{http://www.openslr.org/17/}.} corpus \cite{SnyderD2015MUSAN}. This exposes the models to up to 251 unique speakers\footnote{From the \texttt{librispeech/train-clean-100} folder.} and 843 unique noise types\footnote{From the \texttt{musan/noise/free-sound} folder.} during training. 40 unseen speakers\footnote{From the \texttt{librispeech/test-clean} folder.} and 87 unseen noise types\footnote{From the \texttt{musan/noise/sound-bible} folder.} are used to test the models. 5\% of the training utterances and noises are set aside for validation to help determine training convergence. 

All experiment audio files use a sampling rate of 16000 Hz. Spectrograms are generated using the STFT with a frame size of 1024 samples and a hop size of 256 samples. Per epoch, for each example in the training batch, the sampler mixes a normalized 1-second snippet of a random training speaker's utterance with a normalized 1-second snippet from a random training noise, chosen with uniform probability. There are 100 mixture signals in a batch. Unlike the training mixtures, test mixtures vary in duration; this gives our models an effective RNN lookback size of 1-second.

We assess the proposed ensemble of specialists methodology across two latent spaces. For the \textit{signal degradation} latent space, we instantiate $K=4$ specialists and generate noisy speech mixtures with specific signal-to-noise (SNR) levels---either -5, 0, 5, or 10 dB---for each of the four specialists. Similarly for the \textit{speaker gender} experiment, there are $K=2$ specialists which see a gender-filtered subset of the training data with uniformly varying input SNR values out of the four above listed. In contrast, the baseline model must generalize to all levels of signal degradation and all speaker genders; its training batches consist of 100 mixed gender 1-second-long mixtures with input SNR uniformly distributed between the four values.

All networks are optimized using the Adam optimizer \cite{KingmaD2015adam} with an initial learning rate of $\eta = 0.001$.
The specialist network uses the additive inverse of the SI-SDR metric (Eq. \ref{eq:sisdr}) between $\hat\bs$ and $\bs$ as the loss function, whereas the gating network minimizes the binary cross entropy (BCE) metric between its output, softmax vector $\bp$, and a ground-truth one-hot vector representing the index of the best-suited specialist. Each network variant is trained for approximately three hours on a NVIDIA Titan Xp GPU, after which the validation metric is considered to have converged.

\begin{figure*}[ht]
    \centering
    \begin{subfigure}[t]{\columnwidth}
        \centering
        \includegraphics[width=\columnwidth]{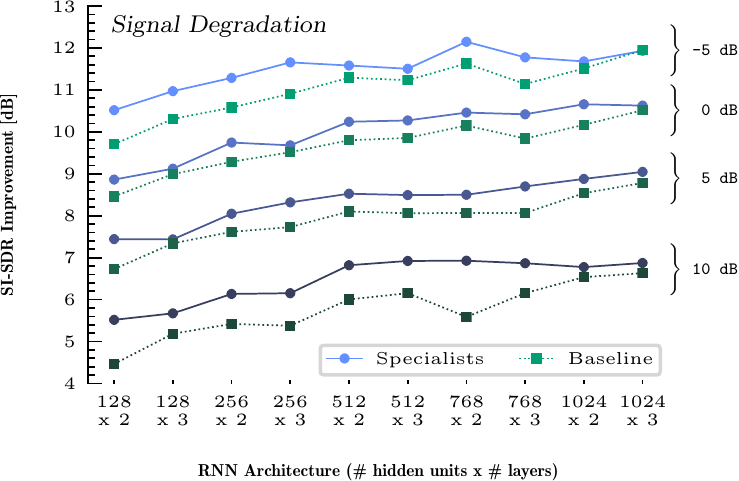}
        \caption{}
        \label{fig:svb}
    \end{subfigure}%
    \hfill
    \begin{subfigure}[t]{\columnwidth}
        \centering
        \includegraphics[width=\columnwidth]{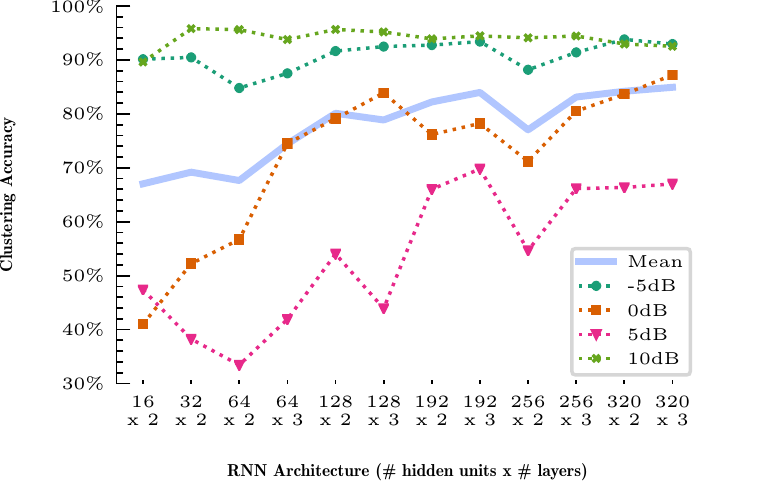}
        \caption{}
        \label{fig:gating_snr}
    \end{subfigure}
    \par\bigskip
    \begin{subfigure}[t]{\columnwidth}
        \centering
        \includegraphics[width=\columnwidth]{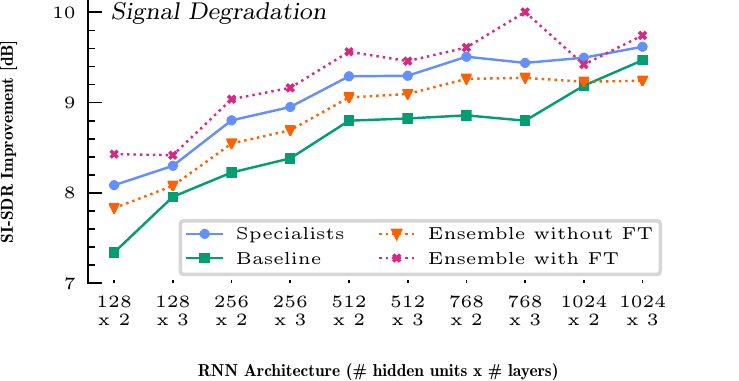}
        \caption{}
        \label{fig:svb_mean_snr}
    \end{subfigure}
    \hfill
    \begin{subfigure}[t]{\columnwidth}
        \centering
        \includegraphics[width=\columnwidth]{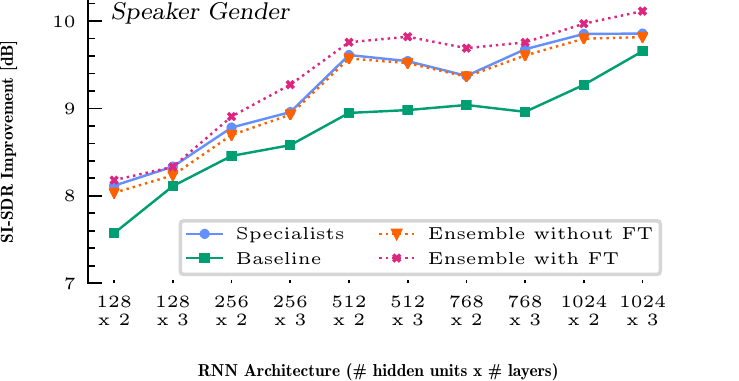}
        \caption{}
        \label{fig:svb_mean_gender}
    \end{subfigure}
    \caption{Results from the signal degradation and speaker gender experiments. The LSTM component of the specialist network increases in computational complexity going across the x-axis on all subplots.}
\end{figure*}

\section{Experiment Results}

We report the denoised signal SI-SDR improvement for all models averaged across 1000 test set mixtures. Figure \ref{fig:svb} compares the test signal speech denoising performance between the four \emph{signal degradation}-based specialists and the one baseline model. It is evident that, at all mixture SNR levels, a neural network specifically trained to denoise those mixtures can outperform a generalist network. This gap in performance is most prominent with the extrema mixture levels (i.e. the -5dB and +10dB mixture SNR cases). As the number of RNN hidden units and layers increases, the performance gap between specialists and baseline model diminishes. With larger network complexity, the generalist's performance eventually matches the specialist's, which saturates after a particular network size.

The specialist curves in Figure \ref{fig:svb} set a theoretical upper bound to the na\"ive ensemble model: even with a perfect gating network, the na\"ive ensemble cannot outperform the sum of its parts. The superior performance of the na\"ive ensemble model to the baseline comes from the fact that each specialist focuses on the smaller subset of the original problem with the same model capacity. It also means that the ensemble of smaller specialist networks can achieve a similar performance to that of the baseline, if the model knows which specialist is the best-suited one. 

Hence, the gating network's classification accuracy matters. The performance of the \emph{signal degradation}-based gating network is shown in Figure \ref{fig:gating_snr}. With a smaller RNN architecture, the gating network can only distinguish the extrema mixture levels with high confidence. Increasing the number of hidden units and layers brings up the classification accuracy of the non-extrema mixture levels (i.e. 0dB and +5dB mixture SNR). Based on these results, we select the $128\times 2$ gating network architecture to be used for the subsequent ensemble experiments, as it adequately clusters test mixtures (with $\approx 80 \%$ accuracy on average) while only incurring a small computational overhead.

Figure \ref{fig:svb_mean_snr} compares the averaged denoising performance of the individual specialists, the baseline, and the ensemble models (with and without fine-tuning) across all four mixture SNR cases. We can see that the na\"ive ensemble improves upon the baseline with a significant margin, but cannot pass the theoretical upper bound set by the oracle choice of specialist. Still, the na\"ive ensemble model can compete as an efficient inference model with the high-complexity baseline model of size $1024\times 2$ with a simpler architectural choice, $512\times 2$. 

Figure \ref{fig:svb_mean_snr} also shows that the fine-tuning step greatly improves our ensemble model, surpassing the oracle specialist upper bound. This suggests that through fine-tuning, the specialists learn to compensate for imperfect classification results from the gating module. We can see that a fine-tuned ensemble with a smaller specialist RNN architecture, $512\times 2$, outperforms the most complex baseline model of size $1024\times 3$. This is a significant amount of computation reduced during the test time, even considering the overhead cost of the $128\times 2$ gating network. 

A similar trend is present in the \emph{speaker gender} experiment, summarized in Figure \ref{fig:svb_mean_gender}. Since this setup consists of only two specialists, the gating network's job is an easier binary classification. A $16\times 2$ RNN architecture sufficiently classifies speaker gender with 90\% classification accuracy. Using that, the na\"ive ensemble achieves near-optimal performance, reaching the upper bound in nearly every architecture. The fine-tuning process lifts the performance even further. 

\section{Conclusion}

In this paper, we showed that neural networks for speech denoising can benefit from the MLE design philosophy, improving performance as well as reducing computational complexity. The specialist networks in our experiment are trained on specific partitions of a large noisy speech corpus across two latent spaces: \emph{signal degradation} and \emph{speaker gender}. By adding the small overhead cost of a gating network, trained to select the best specialist for an input signal, a na\"ive ensemble network is able to reach the theoretical upper bound of all the specialist networks. Furthermore, fine-tuning the ensemble with the inclusion of a sparsity parameter helps the model exceed this theoretical upper bound.

% We envision many possible extensions of the MLE design towards speech denoising. For example, we consider modifying the gating network to output cluster probabilities $\bp$ on a frame-by-frame basis, effectively allowing the proposed model to perform real-time processing. Such a modification might diminish the gains of reduced computational complexity which the originally proposed off-line ensemble model provides; in the worst case scenario, switching specialists with every input frame may require all specialists to be actively involved in the inference. We hope to explore this idea and others while upholding the benefits of sparse specialist activation.

Denoised speech examples and source code for this project are available online at \url{http://saige.sice.indiana.edu/research-projects/sparse-mle}.

% Generated by IEEEtran.bst, version: 1.13 (2008/09/30)

\end{document}